\documentclass[preprintnumbers,a4paper,twocolumn,floats,aps,nofootinbib]{revtex4-1}
\usepackage{graphicx}
\usepackage{amsmath,amssymb,amsfonts}
\usepackage{bbm}
\usepackage{xspace}
\usepackage{verbatim}
\usepackage{bm}
\usepackage{color}
\usepackage{cancel}
\usepackage{ulem}
\usepackage{bbm}
\usepackage[capitalise, english]{cleveref}
\usepackage[separate-uncertainty=true]{siunitx}
\usepackage{tikz}
\usepackage{soul}

\definecolor{tobycolour}{rgb}{.5,.0,.5}
\definecolor{herbicolour}{rgb}{1.,.0,.0}

\definecolor{mkgreen}{rgb}{0.2,.70,.3}

\DeclareSIUnit\parsec{pc}

\usetikzlibrary{patterns,decorations.pathreplacing}
\usetikzlibrary{decorations.pathmorphing}
\usetikzlibrary{decorations.markings}
\usetikzlibrary{arrows,shapes}
\usetikzlibrary{shapes.misc}
\tikzset{
    gluon/.style={decorate, draw=black,
        decoration={coil,amplitude=4pt, segment length=4pt,aspect=0.7}} 
}
\tikzset{
    photon/.style={decorate, decoration={snake}},
}

\makeatletter
\providecommand*{\diff}%
	{\@ifnextchar^{\DIfF}{\DIfF^{}}}
\def\DIfF^#1{%
	\mathop{\mathrm{\mathstrut d}}%
		\nolimits^{#1}\gobblespace}
\def\gobblespace{%
	\futurelet\diffarg\opspace}
\def\opspace{%
	\let\DiffSpace\!%
	\ifx\diffarg(%
		\let\DiffSpace\relax
	\else
		\ifx\diffarg[%
			\let\DiffSpace\relax
		\else
  			\ifx\diffarg\{%
				\let\DiffSpace\relax
			\fi\fi\fi\DiffSpace}

\begin{document}

\title{Validity of the CMSSM interpretation of the diphoton excess}

\preprint{CERN-TH-2016-145, Bonn-TH-2016-05}

\author{Herbi K. Dreiner$^1$}
\email{dreiner@uni-bonn.de}
\author{Manuel E. Krauss$^1$}
\email{mkrauss@th.physik.uni-bonn.de}
\author{Ben O'Leary}
\email{benjamin.oleary@gmail.com}
\author{Toby Opferkuch$^1$}
\email{toby@th.physik.uni-bonn.de}
\author{Florian Staub$^2$}
\email{florian.staub@cern.ch}

\affiliation{$^1$Bethe Center for Theoretical Physics \& Physikalisches Institut der 
Universit\"at Bonn, \\
Nu{\ss}allee 12, 53115 Bonn, Germany}
\affiliation{$^2$Theory Department, CERN, 1211 Geneva 23, Switzerland}

\begin{abstract} 
It has been proposed that the observed diphoton excess at 750~GeV could be explained within the constrained minimal supersymmetric standard model 
via resonantly produced stop bound states. We reanalyze this scenario critically and 
extend previous work to include the constraints from the stability of the electroweak vacuum and 
from the decays of the stoponium into a pair of Higgs bosons. It is shown that the interesting regions of parameter 
space with a light stop and Higgs of the desired mass are ruled out by these constraints. 
This conclusion is not affected by the presence of the bound states because the binding energy is usually very small in 
the regions of parameter space which can explain the Higgs mass. Thus, this also leads to strong constraints on the 
diphoton production cross section which is in general too small. 
\end{abstract}

\maketitle

\section{Introduction}
The diphoton excess seen at the LHC at 750~GeV in the first data set of the 13~TeV run \cite{ATLAS:2015,CMS:2015dxe}  has triggered a lot of excitement.\footnote{The data collected in 2016 \cite{ATLAS:2016eeo,CMS:2016crm} do not confirm this excess. Combining the 2015 and 2016 data sets, the local significance is reduced from $\sim 3-4\,\sigma$ to no more than $\sim 2\,\sigma$ at both ATLAS and CMS.} Many 
different explanations for this excess have been proposed. In weakly coupled theories usually a new fundamental scalar with 
a mass of 750~GeV is introduced to explain this excess; see, for instance, Ref.~\cite{Staub:2016dxq} for an overview. One 
alternative possibility in weakly coupled theories was pointed out in Ref.~\cite{Kats:2016kuz}: it was shown that bound states 
of a pair of colored scalars or fermions with masses of about 375~GeV can explain this excess while being in agreement 
with all other constraints from direct searches. Reference~\cite{Kats:2016kuz} finds that the new  particles should have charge 4/3 
or 5/3 to have a sufficiently large diphoton production cross section. This is based on the assumption that the binding 
energies are small enough for the relativistic calculations to hold. In Ref.~\cite{Choudhury:2016jbc} 
it was claimed that the same idea works in the constrained version of the minimal supersymmetric standard model (CMSSM): 
the bound states are formed by a pair of scalar top partners (stops). In order to be in agreement with the production rate a large 
binding energy was assumed, which causes a large uncertainty on the production cross section.  

The CMSSM is experimentally already extremely challenged, if not excluded, when including the constraints for $(g-2)_\mu$ \cite{Bechtle:2015nua}.
In the perturbatively calculable regions of the CMSSM, it is well known that light stops can no longer be obtained  when including all 
existing constraints. The main reason for this is the Higgs mass which is bounded from above at the tree level and which needs large 
radiative corrections, mainly from stops. This is only possible in the case of one light stop eigenstate if a large mass splitting in the stop 
sector is present. This large splitting is severely constrained by bounds from vacuum stability: if the trilinear coupling is responsible for 
enhancing the Higgs mass and for splitting the two stops, minima in the scalar potential can appear where charge and color get broken 
via vacuum expectation values (VEVs) of the stops. Therefore, we critically  reanalyze the possibility of explaining the diphoton excess within the CMSSM when including these constraints. 

In this context we also comment on the possibility of obtaining very large binding energies of the 
stoponium which might render both perturbative Higgs mass calculations, as well as standard checks of the vacuum stability 
inappropriate. Even if it is questionable that the changes in the Higgs mass would be so  dramatic to 
be in agreement with the measurements, there is an even stronger argument to rule out these parameter regions: the branching ratio of 
the stoponium into a pair of Higgs bosons would be much larger than  into a pair of photons. 

This article is organised as follows: in Sec.~\ref{sec:CMSSM} we briefly review the vacuum stability constraints in the CMSSM, in 
Sec.~\ref{sec:stoponium} we discuss the possible impact  of stoponium bound states, and in Sec.~\ref{sec:num} we show 
our main results based on a  numerical check of the vacuum stability in the CMSSM parameter space with a 375~GeV stop. We 
conclude in Sec.~\ref{sec:conclusion}.

\section{Vacuum stability constraints on a light stop in the CMSSM}
\label{sec:CMSSM}
\subsection{The stop mass and vacuum stability}
The vacuum stability of the CMSSM was studied in detail in Ref.~\cite{Camargo-Molina:2013sta}. It was 
 found that the minimal stop mass 
 which can be 
 obtained in the CMSSM and which has a stable electroweak (EW) vacuum is about 600~GeV when setting $M_{1/2} = 500$~GeV. 
 This value is now already in conflict with the limits from gluino searches, i.e. the lower bound on the stop mass is even 
 higher. In addition, this limit does not include the constraint from the Higgs mass. When adding this constraint, it was found in 
 Ref.~\cite{Chamoun:2014eda} that the minimal stop mass in the CMSSM with a stable vacuum is already above 800~GeV. 

\subsection{The tunneling time}
The limits quoted so far on the stop mass  only  checked if there is a  minimum with 
nonvanishing stop VEVs which is deeper than the EW one. Of course, there is the possibility that the EW vacuum 
is metastable but longlived on cosmological time scales. The usual expression for the decay rate $\Gamma$ per unit volume 
for a false vacuum is given in \cite{Coleman:1977py, Callan:1977pt} as
\begin{equation}
 \Gamma / \text{vol.} = A e^{( -B / \hbar)} ( 1 + \mathcal{O}( \hbar ) )
\label{eq:tunneling_time}
\end{equation}
where $A$ is a factor which depends on eigenvalues of a functional determinant  and $B$ is the bounce action. The $A$ factor is 
typically estimated on dimensional grounds, as it is very complicated to calculate and, because of the  exponentiation of $B$, is 
far less important than getting the bounce action as  accurate as possible.  $A$ is usually taken of order  the renormalization 
scale, and one can feel free to assign an uncertainty of 1 order of magnitude which would change the lifetime by 4 orders. 
However, as we will see, our conclusion about the validity of the proposed scenario does not depend on this. 

$B$ is usually calculated numerically. The most widely used tool for doing this is {\tt CosmoTransitions} 
\cite{Wainwright:2011kj}. In this context one has to keep in mind several effects which could alter the lifetime as calculated with 
{\tt CosmoTransitions}:
\begin{itemize}
 \item[(i)] It is not guaranteed that {\tt CosmoTransitions} always finds the optimal path for tunneling.
 \item[(ii)] There might be other directions in the VEVs when including more  scalar fields beyond 
  the Higgs doublets and stops, which could cause a faster decay of the EW vacuum \cite{Camargo-Molina:2013sta}; for a recent discussion, see also Ref.~\cite{Hollik:2016dcm}.
 \item[(iii)] The inclusion of thermal effects can 
 reduce the likelihood that the Universe is still in a metastable but long-lived vacuum
 \cite{Camargo-Molina:2014pwa}.
 \item[(iv)] Planck suppressed operators can cause a decrease in the lifetime of the EW vacuum \cite{Branchina:2014rva,Lalak:2014qua}.
\end{itemize}
All of these effects can only {\it decrease} the lifetime of the EW vacuum if it is metastable; thus the resulting limits
are conservative.

\subsection{The vacuum lifetime for very light stops}
\begin{figure}[tbh]
\includegraphics[width=\linewidth]{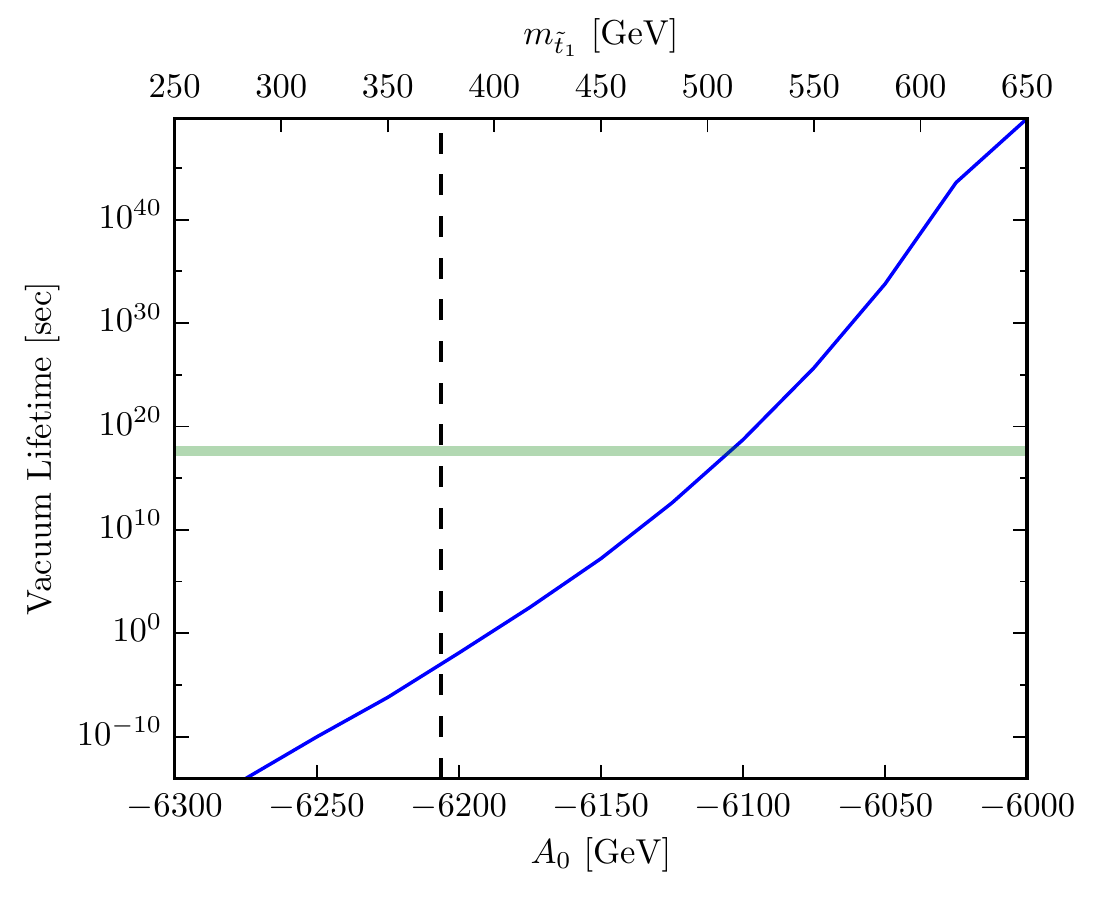}
\caption{The lifetime of the EW vacuum as a function of $A_0$. The other CMSSM parameters are chosen as $m_0=2750$~GeV, $M_{1/2}=750$~GeV, $\tan\beta=15$, and $\mu>0$. Also the mass of the light stop is shown. The black dashed line corresponds to $m_{\tilde t_1} = 375$~GeV while the green  line corresponds to a lifetime of 13.8 billion years.}
\label{fig:lifetime}
\end{figure}

We show in Fig.~\ref{fig:lifetime}  for an example CMSSM parameter point how quickly the 
EW vacuum lifetime decreases with increasing $|A_0|$ and decreasing stop mass. This point has a stable EW vacuum 
for $A_0 \sim 5500$~GeV which corresponds to a stop mass of 850~GeV. Using as a condition to have a lifetime of the 
EW vacuum longer than the current age of the universe allows one to decrease the stop mass to about 525~GeV. However, the point 
with $m_{\tilde t_1}=375~$GeV has a lifetime of only a fraction of a second and is therefore ruled out beyond doubt. Moreover, we have so far not taken into account 
thermal corrections to the tunneling process. As soon as we do that, all points with a stop mass below 700~GeV have a lifetime 
much smaller than the age of the universe.

\section{Stop bound states}
\label{sec:stoponium}
\subsection{Estimate of the binding energy}
\begin{figure}[tbh]
\includegraphics[width=\linewidth]{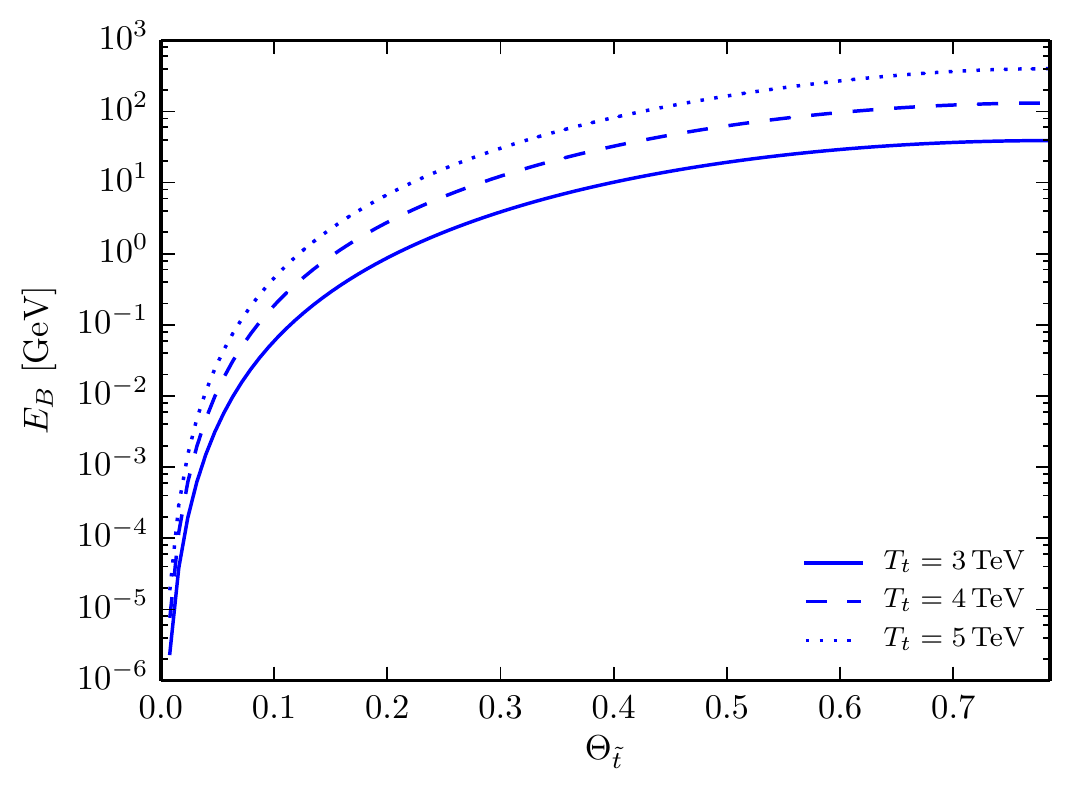}
\caption{The binding energy of two stops with $m_{\tilde t_1}$=375~GeV for $T_t=3$~TeV (solid line), 4~TeV (dashed line), and 5~TeV (dotted line) as a function of the stop mixing angle $\Theta_t$.}
\label{fig:EB}
\end{figure}

It has been pointed out in Ref.~\cite{Giudice:1998dj} that in the case of large trilinear couplings the stops can 
form bound states (``stoponium'', $\sigma_{\tilde t}$) via the exchange of Higgs bosons. A  rough approximation for 
the mass of the bound state was given as
\begin{equation}
M_B = 2 m_{\tilde t_1} \sqrt{1-\frac{1}{(16 \pi)^2 4n^2} \left(\frac{T_t \cos\alpha \sin 2\Theta_{\tilde t}}{\sqrt{2} m_{\tilde t_1} } \right)^4}\,.
\end{equation}
Here, $\Theta_{\tilde t}$ and $\alpha$ are the stop and Higgs mixing angles, respectively, whereas $n$ counts the 
bound state modes. One can see from this equation that two conditions are necessary to have a small mass or a large binding energy which can even be of the order of the EW scale: very large trilinear couplings $T_t$ and a 
large stop mixing $\Theta_{\tilde t}\sim \pi/4$. This was also pointed out in Ref.~\cite{Kang:2016wqi}. The strong dependence on the mixing angle is depicted in Fig.~\ref{fig:EB}. 
Thus, only for mixing above about 0.3 can the binding energy be in the multi-GeV  range, for $T_t$ of 
order a few TeV. That large binding energy would then have some impact on the study of the vacuum stability and one 
might need to take these effects into account. However, for smaller mixing angles, the binding energy is tiny compared to 
the stop mass scale, which is the important scale also for the tunneling processes. 
In these cases one can safely expect that the standard calculations hold. 

\subsection{Correlation between the stoponium binding energy and the light Higgs mass}
\begin{figure*}[hbt]
\centering
 \includegraphics[width=0.49\linewidth]{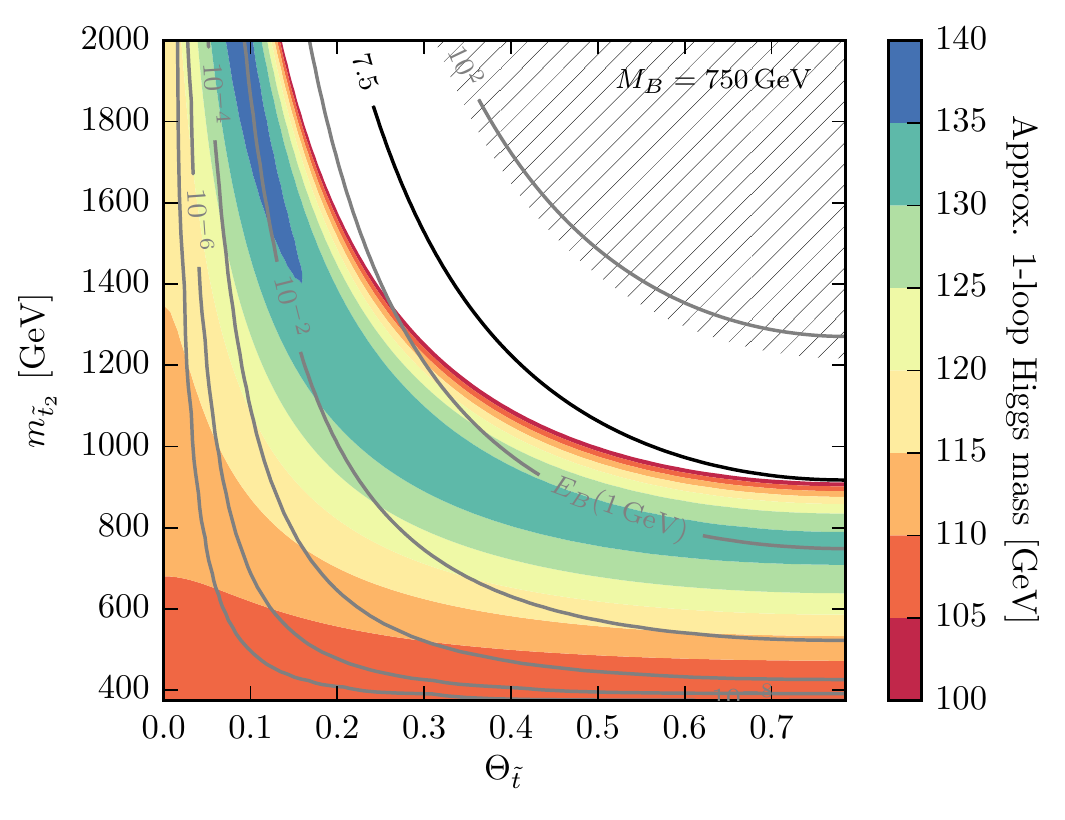}
 \includegraphics[width=0.49\linewidth]{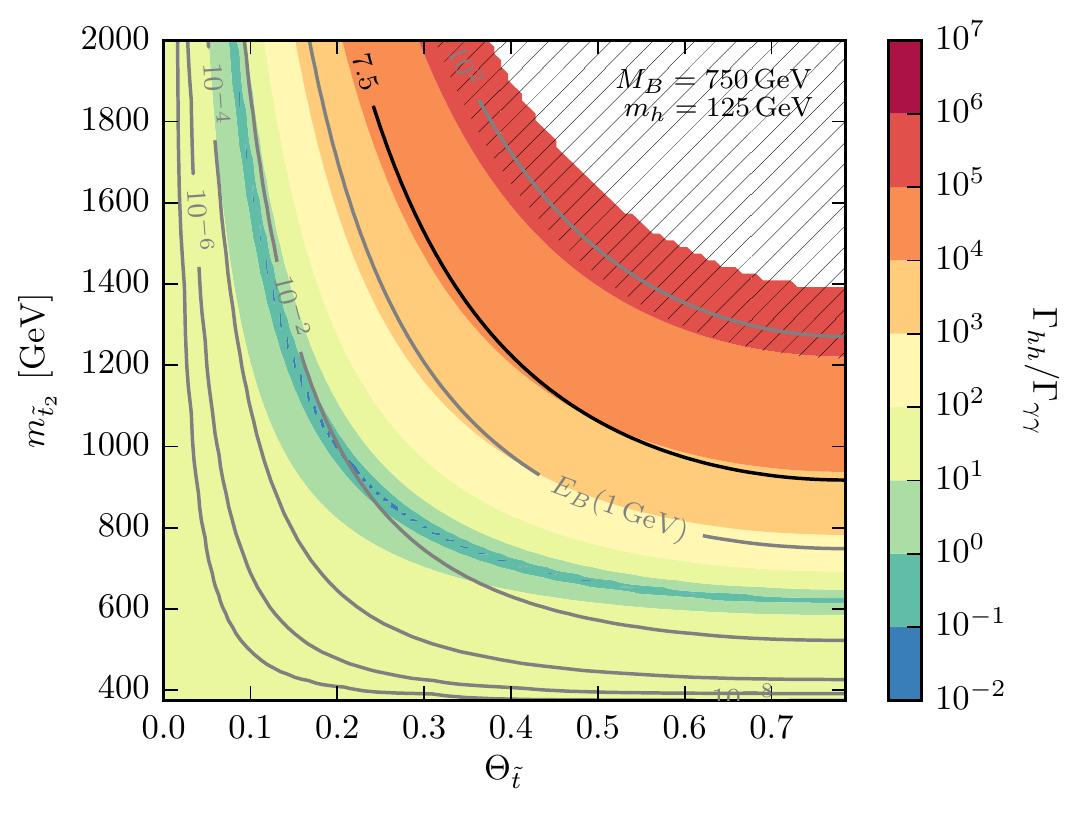}
\caption{Left: the estimate of the one-loop corrected Higgs mass (colored contours) in the $(\Theta_{\tilde t}, 
m_{\tilde t_2})$ plane. Right: the ratio of the partial decay widths of the stoponium into a pair of Higgs bosons (for fixed 
$m_h =125$~GeV)  and a pair of photons. We have set $M_B = 
750$~GeV, respectively. In both figures, the lines are contours of constant $E_B/\text{GeV}$, the black line 
indicates the contour where $E_B$ reaches $10\,\%$ of $M_B$. }
\label{fig:H1loopBinding}
\end{figure*}

We can make a rough estimate to see if the binding energy in the parameter space of interest is expected to be large. For this 
purpose we assume the stop mixing matrix at tree level to be parametrized by
\begin{equation}
m_{\tilde t}^2 = \left(\begin{array}{cc}  m_{LL}^2 & m_t X_t \\  m_t X_t^* & m_{RR}^2 \end{array} \right)\,,
\end{equation}
where $m_{LL}^2$, $m_{RR}^2$ are the sums of soft supersymmetry-breaking F- and D-terms as well as $X_t \equiv T_t/Y_t - \mu 
\cot\beta$. For $\tan\beta \gg 1$ the first term dominates, and we assume this limit for the following brief discussion. 
We will always refer to mass-ordered eigenstates, $\tilde t_{1(2)}$ being the lighter (heavier) stop eigenstate.
Together 
with the well-known expression for the one-loop corrections to the Higgs mass via (s)tops in the decoupling limit $M_A \gg M_Z$ 
\cite{Haber:1993an,Carena:1995wu,Martin:1997ns,Heinemeyer:1998np,Heinemeyer:1999be,Carena:2000dp},
\begin{equation}
 \delta m_h^2 = \frac{3}{2 \pi^2}\frac{m_t^4}{v^2} \left[\log\frac{M_S^2}{m_t^2} + \frac{X_t^2}{M_S^2} 
 \left(1-\frac{X_t^2}{12 M_S^2}\right) \right]\,,
\end{equation}
with $M_S =\sqrt{m_{\tilde t_1} m_{\tilde t_2}}$, one can express the one-loop corrected Higgs mass as a function of 
$m_{\tilde t_2}$ and $\Theta_{\tilde t}$, when fixing the stoponium mass at $M_B = 750$~GeV. 
For the same parameters,  we also compute the binding energy. The combined results are shown in the left 
panel of Fig.~\ref{fig:H1loopBinding}.
One can see that for light stop masses, $m_{{\tilde t}_2}$, the maximal enhancement for the Higgs mass appears, as 
expected, for maximal stop mixing. However, for heavier stops, a smaller stop mixing is preferred. One can also see that, in the
interesting region with the largest corrections to the Higgs mass, the stoponium binding energy is usually small and often even 
below the GeV range. Very large binding energies of order the EW scale only appear in parameter regions in which the 
light Higgs would even become tachyonic, because of huge negative one-loop corrections. Thus, in general one can assume 
that for parameter regions which lead to the correct Higgs mass using the standard calculations, also the standard checks for 
the vacuum stability do indeed hold. Moreover, the production rate can be calculated using the expressions for a pure QCD 
bound state; it turns out to be too small to explain the observed diphoton excess.

\subsection{Strongly coupled stoponium: Di-Higgs decays}
These arguments are valid as long as we are situated in a ``normal'' environment, where perturbative calculations hold and the 
Higgs boson is a pure elementary particle. Even in the setup of the MSSM, however, it might be possible to find regions with 
very large stoponium binding energies. One might argue that one 
cannot trust perturbative evaluations of the Higgs mass in the regions where it gets tachyonic due to the large trilinear couplings 
involved and that instead lattice calculations would be more appropriate for a calculation of $m_h$.\footnote{This was actually 
{\it not} the ansatz of Ref.~\cite{Choudhury:2016jbc}, which made use of the standard Higgs mass calculations.} 
In Fig.~\ref{fig:H1loopBinding}, we hatch the regions where the perturbative calculation can no longer be trusted, where we have conservatively taken this region to begin at  $E_B = 0.1 M_B$.
Let us assume for the time being that we are in this strongly coupled phase and that a $125~$GeV Higgs mass is possible with large $E_B$ due to Higgs exchange. This immediately raises the question how important the decay of the stoponium into a 
pair of Higgs bosons becomes with respect to the desired diphoton decay, as these regions feature very large trilinear 
couplings. 
Fortunately, all partial widths scale in the same way with the wave function at the origin (and therefore with the binding energy); 
therefore, this factor drops out when calculating ratios so that solid predictions can be made. The respective formulas can, for 
instance, be taken from Ref.~\cite{Martin:2008sv}; for earlier work see also \cite{Drees:1993uw}. In the right-hand panel of 
Fig.~\ref{fig:H1loopBinding} we show the resulting ratio of $\Gamma_{hh}/\Gamma_{\gamma \gamma}$ in the same plane  as 
in the left-hand panel. Following the outlined argumentation, we assume here that the perturbative computation does not reproduce the correct Higgs mass because of the large trilinear couplings involved. Therefore, we fix the physical Higgs mass to 125~GeV in the computation of the branching ratio in order to obtain the correct decay kinematics in the entire plane. As expected, the di-Higgs 
decay rate is much larger than the diphoton decay rate in the regions where the binding energy due to Higgs exchange becomes 
large, i.e., in regions with large $|X_t|$. Interestingly, all of the parameter space which features a tachyonic Higgs, assuming 
standard perturbative calculations, has a ratio of the partial widths $\Gamma_{hh}/\Gamma_{\gamma \gamma}$ larger than $10^3$. For regions where the binding energy reaches a percent 
of the bound state mass the partial width ratio is larger than $10^4$. 

Let us now compare the results to the experiment. In Ref.~\cite{ATLAS:2014rxa}, a search for
resonant Higgs pair production in the $b\bar b b \bar b$ final state was performed, setting limits of $\sim 12~$fb at $\sqrt s=8~$TeV. Assuming 
gluon fusion for the production mechanism of the stoponium and taking the most conservative best-fit value for the necessary diphoton cross section at $\sqrt{s}=13~$TeV from Ref.~\cite{Nilles:2016bjl}, 
we arrive at the experimental bound $\Gamma_{hh}/\Gamma_{\gamma \gamma}<64$.\footnote{Other reference values for the 
diphoton cross section (see, for instance, Ref.~\cite{Buckley:2016mbr}) suggest even more constrained ratios of  $\Gamma_{hh}/\Gamma_{\gamma \gamma}<42$.}
This eliminates all of the parameter space where the stoponium has large binding energy, in clear contradiction to Ref.~\cite{Choudhury:2016jbc}. 

\subsection{Stoponium-Higgs mixing}
Thus far, we have assumed that the 125~GeV Higgs is an elementary particle and solely a mixture of the MSSM fields 
$H_u$ and $H_d$. We drop this assumption now in order to see if this would alter the conclusions of the previous sections. 
The situation becomes more complicated as soon as the stoponium mixes with the Higgs and takes part in electroweak 
symmetry breaking through the acquisition of a VEV. Although this situation is highly disfavored, given the almost perfect 
agreement of the Higgs signal strength and coupling measurements with the Standard Model, let us assume for the sake 
of argument that this situation is possible. In this case, the stoponium is a new scalar degree of freedom in the theory, introducing 
a new direction in the scalar potential which is not calculable using perturbative methods. Such a scenario arises for very large 
values of $|X_t/M_S|$ \cite{Cornwall:2012ea}, corresponding to such a tightly bound state that a lattice calculation is the only 
reliable technique, as of today. Unfortunately, these calculations have not yet been performed. However, for our purposes such 
precise predictions are not necessary as the rough order of magnitude of the relative partial decay widths into $hh$ and 
$\gamma \gamma$ will not be affected. In particular, the rate into a pair of light Higgs bosons would still be huge due to the 
necessarily large $|X_t|$ values. A conservative estimate would be to parametrize the two scalar states at 125 and 750~GeV as 
$\Phi_{125} = h \cos \phi + \sigma_{\tilde t} \sin \phi$ and $\Phi_{750} = - h \sin \phi + \sigma_{\tilde t} \cos \phi$. Projecting out 
only the $\sigma_{\tilde t}$ production followed by the $\sigma_{\tilde t} \to hh$ decay, the respective partial width roughly 
scales  as $\cos^6 \phi$, leading to a suppression of $\frac{1}{8}$ in the case of maximal mixing.\footnote{This is a very 
conservative estimate as it only corresponds to the decay shown in the figure, neglecting all contributions from the $h \to 
\sigma_{\tilde t} \sigma_{\tilde t}$ projection as well as the mixed projections.} Even 
allowing for another order of magnitude uncertainty due to the now undetermined VEV structure cannot rescue the points with 
$E_B/M_B > 10^{-2}$ from being experimentally excluded. As a result, the regions in parameter space which could potentially 
feature a stoponium condensate participating at EW symmetry breaking, i.e., those regions with binding energies of the order 
of the EW scale, are excluded by many orders of magnitude, far beyond any imaginable source of uncertainty for the ratio 
$\Gamma_{hh}/\Gamma_{\gamma \gamma}$. 

The entire discussion has so far neglected the high-scale boundary conditions present in the CMSSM. We find that in 
regions of parameter space where $|X_t/M_S| \gtrsim 15$, which is the range where the critical coupling for EW symmetry 
breaking through stop condensates is reached \cite{Cornwall:2012ea}, the lightest stau always becomes tachyonic.\footnote{This assumes the standard renormalization group equation
evaluation, but is also rather robust 
against deviations in the SM Yukawa couplings, which could be caused by the stoponium-Higgs mixing.} 

\subsection{The possible stoponium binding energy}
We can briefly summarize our discussion of the possible stoponium binding energy. Very large binding energies are 
immediately ruled out by the di-Higgs decay rate. Consequently, the maximal binding energy of a 750~GeV stoponium from 
pure Higgs exchange is less than $1\,\%$. Therefore, it is of the order of, or smaller than, the typical binding energy from 
perturbative QCD \cite{Drees:1993uw}. Such a QCD bound state, as has been discussed in Ref.~\cite{Choudhury:2016jbc}, 
is insufficient to produce the required diphoton rate. Moreover, these small binding energies render the usual vacuum stability 
considerations, which involve much higher scales, fully consistent.

\section{The CMSSM with a 375~GeV stop}
\label{sec:num}
Even though this scenario is already highly disfavored by a production rate which is too small, we nevertheless discuss the 
impact of vacuum stability constraints in more detail. 
The reason is that the calculation of the production
cross section still includes uncertainties,
potentially increasing 
the production cross section such that the resulting signal is consistent to within $2\,\sigma$ of the observed 
excess. 
In addition, one might consider the case that there are other contributions
to the diphoton rate in the CMSSM-like sbottomium in the large $\tan\beta$
limit. Here, we aim to exclude light stoponium bound states within the CMSSM, independently of the 
diphoton cross section, leaving the results also applicable in the currently more realistic case that the diphoton excess turns out to be a 
statistical fluctuation.
Therefore, we perform a  numerical analysis of the CMSSM in the remaining parameter space with $M_B\simeq 2 
m_{\tilde t_1}$, where, in contrast to Fig.~\ref{fig:lifetime}, the standard calculations reproduce the observed Higgs 
mass and the vacuum should be sufficiently stable. 
The results we find  also apply
to the CMSSM
parameter space with light stops in that ballpark which do not necessarily form
bound states.
For this purpose we use a {\tt SARAH} 
\cite{Staub:2008uz,Staub:2009bi,Staub:2010jh,Staub:2012pb,Staub:2013tta,Staub:2015kfa} generated {\tt SPheno} 
\cite{Porod:2003um,Porod:2011nf} version to calculate the mass spectrum, including the full one-loop corrections to the stops and 
the dominant two-loop corrections to the Higgs states \cite{Goodsell:2014bna,Goodsell:2015ira}. To check the vacuum stability we 
use {\tt Vevacious} \cite{Camargo-Molina:2013qva}. {\tt Vevacious} finds the global minimum of the one-loop corrected effective 
potential and calls {\tt CosmoTransitions} to calculate the lifetime if necessary. For our checks we used {\tt Vevacious} with the 
model files for the MSSM with real VEVs for the neutral Higgs doublets and the two stops which were also generated by {\tt SARAH}.

\subsection{Scan of the ($m_0$,$M_{1/2}$) plane}
\begin{figure}[hbt]
 \includegraphics[width=\linewidth]{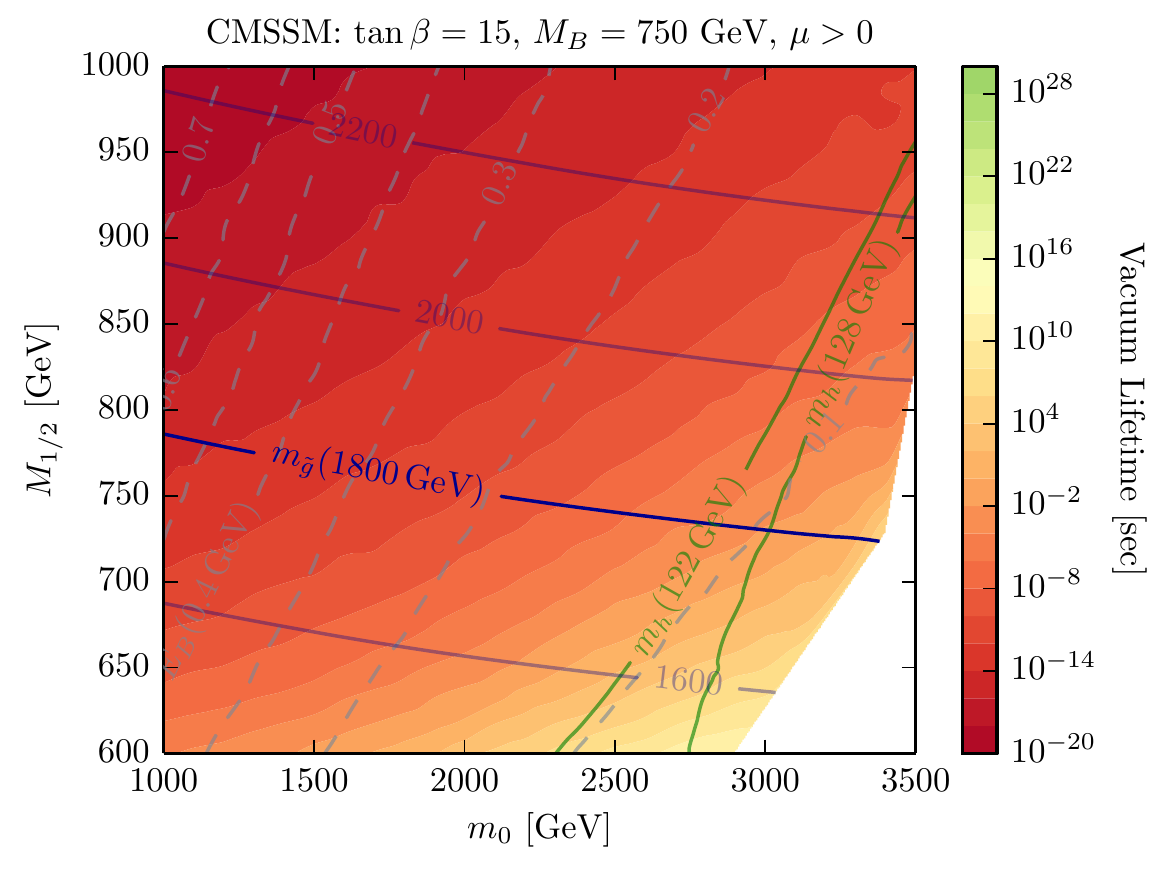}
\caption{The lifetime of the EW vacuum in seconds (colored contours) as calculated with the combination 
{\tt SPheno}-{\tt Vevacious}-{\tt CosmoTransitions}. For comparison, the age of the universe is $\sim 4.35\times 10^{17}~$s. 
Between the two green contour lines, the Higgs mass lies in the range $125\pm3$~GeV. The stoponium binding energy from Higgs exchange in GeV 
is shown as grey dashed contours, while the dark blue contours indicate the gluino mass in GeV. The line for $m_{\tilde g}=1800\,$GeV is highlighted, as it roughly corresponds to the current lower experimental bound on the gluino mass. 
}
\label{fig:m0M12}
\end{figure}

In order to check if it is possible to have a 375~GeV stop in the CMSSM with the correct Higgs mass we perform a scan in the 
following ranges:
\begin{eqnarray*}
&m_0 = [1,3.5]~\text{TeV}, \hspace{1cm} M_{1/2}=[0.6,1.0]~\text{TeV},
\end{eqnarray*}
where we fix both $\tan\beta=15$ and $\mu>0$. $A_0$ is fit at each point to ensure a light stop at 375~GeV. For  
larger $M_{1/2}$ values it is not possible to find points with the correct Higgs mass without going to even larger $m_0$ values. 
However, if $m_0$ is too large the performed fixed order calculation of the Higgs mass suffers from a large uncertainty. We therefore 
restrict ourself to values within this range. It turns out that the lifetime of the EW vacuum quickly drops with 
increasing $M_{1/2}$; i.e., this restriction does not affect the generality of our results. One also expects that the results are robust 
against changes of $\tan\beta$. However, for very large $\tan\beta$ the vacuum stability issue becomes more severe, due to the 
possible appearance of stau and sbottom VEVs. The results are summarized in Fig.~\ref{fig:m0M12}. 

One can see that the entire region of the parameter space which is consistent with the Higgs mass measurements and 
accommodates a light stop has a metastable vacuum. In this range, the lifetime of the metastable vacuum is 
always very short on cosmological time scales. Moreover, the region with small $M_{1/2}$ where the lifetime exceeds 
one second is in conflict with the current limits from gluino searches which exclude masses up to $\sim 1.8~$TeV (see, e.g., Refs.~\cite{Khachatryan:2016xvy,Aad:2016eki} 
for recent LHC Run-II results). 
Thus, even when assigning a generous theoretical 
error to the lifetime calculation, the conclusion does not change. Additionally for low $M_{1/2}$, one can see that 
the binding energy of the stoponium lies at most in the 100~MeV range; i.e., this effect cannot have an impact on the 
validity of the vacuum stability results. Finally, because of this small binding energy, the standard calculation of the 
cross section of  stoponium production, followed by the decay to two photons, is 
also expected to be valid in the entire plane. This results in an 
insufficient diphoton cross section for the entire $(m_0, M_{1/2})$ plane. 

\subsection{Maximal binding energy}
\label{sec:MaxEB}
\begin{figure*}[hbt]
 \includegraphics[width=\linewidth]{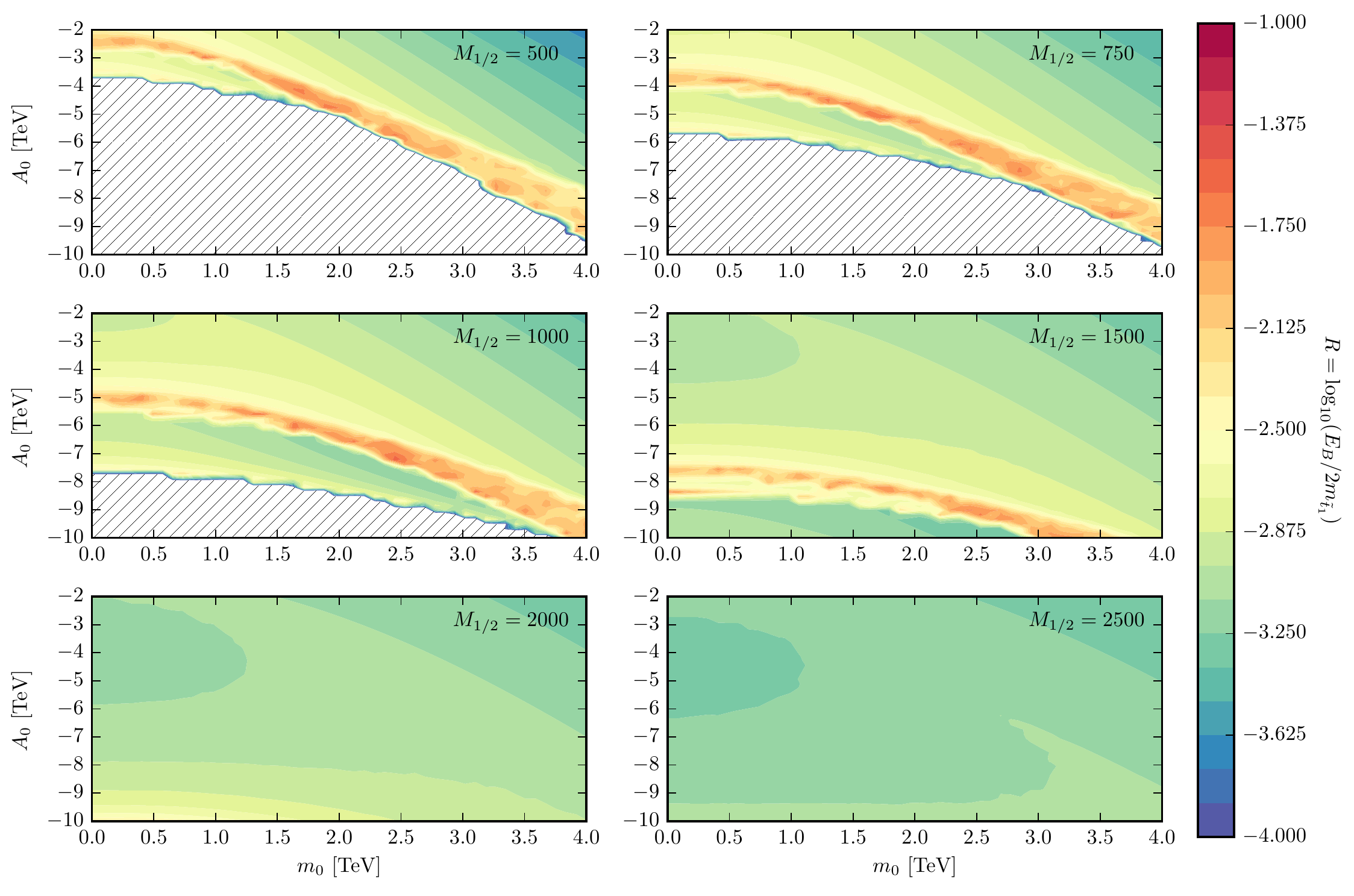}
\caption{The logarithm of the ratio $E_B/2m_{\tilde{t}_1}$ as a function of $A_0$ and $m_0$, where $\tan\beta$ is chosen pointwise to maximize the logarithm. Note that each panel represents a different choice of $M_{1/2}$. The hatched regions correspond to parameter space where the constraints $m^2_{\tilde{\tau}_R}>0$, $m^2_{\tilde{t}_R}>0$ 
and $m_{\tilde{t}_1}>0$ are not satisfied.}
\label{fig:maxR}
\end{figure*}

The results of the previous section show that both the binding energies and vacuum lifetimes for particular combinations of CMSSM parameters are small. Here we demonstrate that small binding energies are indeed a generic feature under the assumption of the CMSSM boundary conditions. Shown in Fig.~\ref{fig:maxR} is the logarithm of the ratio $E_B/2m_{\tilde{t}_1}$ as a function of $A_0$ and $m_0$ for fixed values of $M_{1/2}$. For each parameter point, $(m_0,A_0)$, the largest value of $R=\log_{10}(E_B/2m_{\tilde{t}_1})$ is taken as $\tan\beta$ is varied between the values $(2,60)$ including the following constraints: (i) $m^2_{\tilde{\tau}_R}>0$, (ii) $m^2_{\tilde{t}_R}>0$, (iii) $m_{\tilde{t}_1} > 75$~GeV.\footnote{The value of 75~GeV is a rather arbitrary but very conservative choice. More realistic cuts of 175~GeV to circuvment search limits, or even 375~GeV, in order to have a bound state of around 750~GeV lead to an even smaller upper limit of $R$.} Regions where these conditions are not satisfied for any value of $\tan\beta$ correspond to the hatched regions. The results of these show that the binding energy never exceeds $4\%$ of the mass of the bound state. The reason being is that large $A_0$ values are required for sizable binding energies. However, these large $A_0$ values also enter the renormalization group equations and split the stop-left and stop-right soft SUSY-breaking masses. This results in a reduced stop mixing angle.  Finally large $A_0$ values also lead to negative singlet soft masses for the staus and stops in the case of small $M_{1/2}$. In general, one can find for larger $\tan\beta$ values larger binding energies because of a smaller mass splitting between  $m_{\tilde{t}_R}$ and  $m_{\tilde{b}_L}$, i.e., larger stop mixing. However, at some points the staus become tachyonic and prevent a further increase of $\tan\beta$.

\subsection{Proposed benchmark scenarios in literature}
\label{sec:BP}

In Ref.~\cite{Choudhury:2016jbc}, three benchmark points with a light stop were proposed, which are consistent with the Higgs 
mass and the dark matter relic density. 
For completeness, we use our numerical setup to check the stability of these three points as well. The results are summarized 
in Table~\ref{tab:results} and confirm the previous discussion.

\begin{table}[h]
 \renewcommand\arraystretch{1.10}
\begin{tabular}{cccc}
\toprule
                                   & BP1 & BP2 & BP3 \\
\colrule
$m_0$~[GeV]                        & 2855  & 3199 & 3380 \\
$M_{1/2}$~[GeV]                    & 755   & 860 & 910 \\
$\tan\beta$                        &  15   & 15  & 15 \\
$A_0$~[GeV]                              & $-6405^*$ & $-7205^*$ & $-7620^*$ \\
\colrule
$m_{\tilde t_1}$~[GeV]             & 375   & 425 &444 \\ 
$m_{\tilde t_2}$~[GeV]             &2226   &2495 &2632 \\
$m_{\tilde g}$~[GeV]               &1837   & 2070&2181 \\
$m_h$~[GeV]                        &122    & 122 &122 \\
\colrule
$T_t$~[GeV]                        &$-2960$                      &$-3333$               &$-3520$        \\
$\Theta_{\tilde t}$                         & 0.118                     & 0.106                & 0.101             \\
$E_B$~[GeV]                       & 0.108                     & 0.078                & 0.067             \\
\colrule
$V_{\rm EW}$~[GeV${}^4$]               & $-9.8\times 10^7$              & $-9.8\times 10^7$         & $-9.7\times 10^7$    \\ 
$V_{\rm CCB}$~[GeV${}^4$]              & $-3.7\times 10^{12}$           & $-6.4\times 10^{12}$      & $-8.2\times 10^{12}$  \\
$v_d,v_u$ [TeV]                    & 0.9, 2.8                  & 1.1, 3.2             & 1.2, 3.4 \\
$v_{\tilde t_L}, v_{\tilde t_R}$ [TeV]& 2.2, $-3.0$             & $-2.7$, $-3.7$           & 2.7, 3.7 \\
$\tau$~[s]                         & $2.6\times 10^{-4}$       & $5.9\times 10^{-4}$      & $3.9\times 10^{-4}$  \\
\botrule
\end{tabular}
\caption{The benchmark points proposed by Ref.\cite{Choudhury:2016jbc}. $V_{\rm EW}$ is the depth of the EW vacuum, 
$V_{\rm CCB}$ the depth of the global vacuum with the given Higgs and stop VEVs $v_x$ (with $x=d,u,\tilde t_L, \tilde t_R$). 
$\tau$ is the life-time of the EW vacuum. Note: we slightly adjusted the input values of $A_0$ in order to get the same stop 
masses as given in Ref.\cite{Choudhury:2016jbc}. Since {\tt SPheno} uses different matching conditions 
than {\tt suspect} \cite{Djouadi:2002ze} to calculate the top Yukawa coupling, the running stop mass parameters for the same 
input are slightly different. 
Since we had to {\it decrease} $|A_0|$ compared to Ref.\cite{Choudhury:2016jbc}, the life-time using the original values would be even 
shorter. } 
\label{tab:results}
\end{table}

We see that all three points have a global minimum where charge and color are broken via stop VEVs in the TeV 
range. The depth of color breaking minima is 5 to 6 orders of magnitude deeper than the EW vacuum. This also 
explains the very fast decay of the EW vacuum: all three points have a lifetime which is a tiny fraction of a second. Thus, 
one sees that the energy scales which are important in this calculation are several orders of magnitude above the 
binding energy of the stoponium. Moreover,  the calculated tunneling rate is so huge that not even the assumption of a 
large uncertainty on the coefficient $A$ in \cref{eq:tunneling_time} could possibly alter the above conclusion.

\section{Conclusion}
\label{sec:conclusion}
We have critically reviewed the explanation of the diphoton excess via stop bound states in the CMSSM as proposed in 
Ref.~\cite{Choudhury:2016jbc}. We have discussed that stops in the CMSSM with masses of 375~GeV 
cause charge and 
color breaking minima. This is in particular the case when the constraints from the Higgs mass measurement are included. 
We have summarized results in the literature which find that the lower limit on the stop mass in the CMSSM is about 
800~GeV if the electroweak vacuum should be stable. These limits are certainly weaker if the possibility of a 
metastable but long-lived minimum is considered. These conclusions are not affected by the appearance 
of bound state effects because the binding energy in the experimentally allowed parameter region is very small compared to 
the other relevant scales in the calculation. In addition, because of this small binding energy
the cross section to produce the diphoton signal is too small.

The {\it ad hoc} assumption of large binding energies which makes the calculation of the Higgs mass, as well as the checks for the vacuum stability more difficult is also ruled out by the much too large decay rate of the stoponium into a pair of Higgs bosons. 

Taking all these effects into account, it is not possible to explain the diphoton signal in the CMSSM. Whether the general MSSM would survive, given the excess is confirmed, is currently under debate. There is the claim that it might be possible to obtain a sufficient diphoton cross section in fine-tuned parameter regions of the MSSM with a large $\mu$ term \cite{Djouadi:2016oey}. However, this possibility also lacks the proof of existence since so far no valid parameter regions consistent with the constraints from vacuum stability could be presented.

\section*{Acknowledgments}
We thank A. Kusenko and M. Drees for very helpful discussions as well as D. Choudhury and K. Ghosh for sharing their 
thoughts about this topic. M. E. K. is supported
by the DFG Research Unit 2239 ``New Physics at the LHC''.

\bibliography{CMSSM.bib}

\begin{thebibliography}{10}

\bibitem{ATLAS:2015}
ATLAS,
\newblock {\normalfont 2015} , ATLAS-CONF-2015-081.

\bibitem{CMS:2015dxe}
CMS,
\newblock {\normalfont 2015} , CMS-PAS-EXO-15-004.

\bibitem{ATLAS:2016eeo}
ATLAS,
\newblock ATLAS-CONF-2016-059 (2016).

\bibitem{CMS:2016crm}
CMS,
\newblock CMS-PAS-EXO-16-027 (2016).

\bibitem{Staub:2016dxq}
F.~Staub {\em et~al.},
\newblock (2016), 1602.05581.

\bibitem{Kats:2016kuz}
Y.~Kats and M.~J. Strassler,
\newblock JHEP {\bf 05}, 092 (2016), 1602.08819.

\bibitem{Choudhury:2016jbc}
D.~Choudhury and K.~Ghosh,
\newblock (2016), 1605.00013.

\bibitem{Bechtle:2015nua}
P.~Bechtle {\em et~al.},
\newblock Eur. Phys. J. {\bf C76}, 96 (2016), 1508.05951.

\bibitem{Camargo-Molina:2013sta}
J.~E. Camargo-Molina, B.~O'Leary, W.~Porod, and F.~Staub,
\newblock JHEP {\bf 12}, 103 (2013), 1309.7212.

\bibitem{Chamoun:2014eda}
N.~Chamoun, H.~K. Dreiner, F.~Staub, and T.~Stefaniak,
\newblock JHEP {\bf 08}, 142 (2014), 1407.2248.

\bibitem{Coleman:1977py}
S.~R. Coleman,
\newblock Phys. Rev. {\bf D15}, 2929 (1977),
\newblock [Erratum: Phys. Rev.D16,1248(1977)].

\bibitem{Callan:1977pt}
C.~G. Callan, Jr. and S.~R. Coleman,
\newblock Phys. Rev. {\bf D16}, 1762 (1977).

\bibitem{Wainwright:2011kj}
C.~L. Wainwright,
\newblock Comput. Phys. Commun. {\bf 183}, 2006 (2012), 1109.4189.

\bibitem{Hollik:2016dcm}
W.~G. Hollik,
\newblock JHEP {\bf 08}, 126 (2016), 1606.08356.

\bibitem{Camargo-Molina:2014pwa}
J.~E. Camargo-Molina, B.~Garbrecht, B.~O'Leary, W.~Porod, and F.~Staub,
\newblock Phys. Lett. {\bf B737}, 156 (2014), 1405.7376.

\bibitem{Branchina:2014rva}
V.~Branchina, E.~Messina, and M.~Sher,
\newblock Phys. Rev. {\bf D91}, 013003 (2015), 1408.5302.

\bibitem{Lalak:2014qua}
Z.~Lalak, M.~Lewicki, and P.~Olszewski,
\newblock JHEP {\bf 05}, 119 (2014), 1402.3826.

\bibitem{Giudice:1998dj}
G.~F. Giudice and A.~Kusenko,
\newblock Phys. Lett. {\bf B439}, 55 (1998), hep-ph/9805379.

\bibitem{Kang:2016wqi}
Z.~Kang,
\newblock (2016), 1606.01531.

\bibitem{Haber:1993an}
H.~E. Haber and R.~Hempfling,
\newblock Phys. Rev. {\bf D48}, 4280 (1993), hep-ph/9307201.

\bibitem{Carena:1995wu}
M.~Carena, M.~Quiros, and C.~E.~M. Wagner,
\newblock Nucl. Phys. {\bf B461}, 407 (1996), hep-ph/9508343.

\bibitem{Martin:1997ns}
S.~P. Martin,
\newblock (1997), hep-ph/9709356,
\newblock [Adv. Ser. Direct. High Energy Phys.18,1(1998)].

\bibitem{Heinemeyer:1998np}
S.~Heinemeyer, W.~Hollik, and G.~Weiglein,
\newblock Eur. Phys. J. {\bf C9}, 343 (1999), hep-ph/9812472.

\bibitem{Heinemeyer:1999be}
S.~Heinemeyer, W.~Hollik, and G.~Weiglein,
\newblock Phys. Lett. {\bf B455}, 179 (1999), hep-ph/9903404.

\bibitem{Carena:2000dp}
M.~Carena {\em et~al.},
\newblock Nucl. Phys. {\bf B580}, 29 (2000), hep-ph/0001002.

\bibitem{Martin:2008sv}
S.~P. Martin,
\newblock Phys. Rev. {\bf D77}, 075002 (2008), 0801.0237.

\bibitem{Drees:1993uw}
M.~Drees and M.~M. Nojiri,
\newblock Phys. Rev. {\bf D49}, 4595 (1994), hep-ph/9312213.

\bibitem{ATLAS:2014rxa}
ATLAS,
\newblock {\normalfont 2015} , ATLAS-CONF-2014-005.

\bibitem{Nilles:2016bjl}
H.~P. Nilles and M.~W. Winkler,
\newblock JHEP {\bf 05}, 182 (2016), 1604.03598.

\bibitem{Buckley:2016mbr}
M.~R. Buckley,
\newblock Eur. Phys. J. {\bf C76}, 345 (2016), 1601.04751.

\bibitem{Cornwall:2012ea}
J.~M. Cornwall, A.~Kusenko, L.~Pearce, and R.~D. Peccei,
\newblock Phys. Lett. {\bf B718}, 951 (2013), 1210.6433.

\bibitem{Staub:2008uz}
F.~Staub,
\newblock (2008), 0806.0538.

\bibitem{Staub:2009bi}
F.~Staub,
\newblock Comput. Phys. Commun. {\bf 181}, 1077 (2010), 0909.2863.

\bibitem{Staub:2010jh}
F.~Staub,
\newblock Comput. Phys. Commun. {\bf 182}, 808 (2011), 1002.0840.

\bibitem{Staub:2012pb}
F.~Staub,
\newblock Comput. Phys. Commun. {\bf 184}, 1792 (2013), 1207.0906.

\bibitem{Staub:2013tta}
F.~Staub,
\newblock Comput. Phys. Commun. {\bf 185}, 1773 (2014), 1309.7223.

\bibitem{Staub:2015kfa}
F.~Staub,
\newblock Adv. High Energy Phys. {\bf 2015}, 840780 (2015), 1503.04200.

\bibitem{Porod:2003um}
W.~Porod,
\newblock Comput. Phys. Commun. {\bf 153}, 275 (2003), hep-ph/0301101.

\bibitem{Porod:2011nf}
W.~Porod and F.~Staub,
\newblock Comput. Phys. Commun. {\bf 183}, 2458 (2012), 1104.1573.

\bibitem{Goodsell:2014bna}
M.~D. Goodsell, K.~Nickel, and F.~Staub,
\newblock Eur. Phys. J. {\bf C75}, 32 (2015), 1411.0675.

\bibitem{Goodsell:2015ira}
M.~Goodsell, K.~Nickel, and F.~Staub,
\newblock Eur. Phys. J. {\bf C75}, 290 (2015), 1503.03098.

\bibitem{Camargo-Molina:2013qva}
J.~E. Camargo-Molina, B.~O'Leary, W.~Porod, and F.~Staub,
\newblock Eur. Phys. J. {\bf C73}, 2588 (2013), 1307.1477.

\bibitem{Khachatryan:2016xvy}
CMS, V.~Khachatryan {\em et~al.},
\newblock (2016), 1603.04053.

\bibitem{Aad:2016eki}
ATLAS, G.~Aad {\em et~al.},
\newblock Phys. Rev. {\bf D94}, 032003 (2016), 1605.09318.

\bibitem{Djouadi:2002ze}
A.~Djouadi, J.-L. Kneur, and G.~Moultaka,
\newblock Comput. Phys. Commun. {\bf 176}, 426 (2007), hep-ph/0211331.

\bibitem{Djouadi:2016oey}
A.~Djouadi and A.~Pilaftsis,
\newblock (2016), 1605.01040.

\end{thebibliography}
\bibliographystyle{h-physrev}

\end{document}